# High efficient solar evaporation by airing multifunctional textile

Guilong Peng[1,2], Shichen Deng[1,2], Swellam W. Sharshir[1,3,4], Dengke Ma[1,2], A.E. Kabeel[5], Nuo Yang[1,2,*]

1 State Key Laboratory of Coal Combustion, Huazhong University of Science and Technology, Wuhan 430074, China

2 Nano Interface Center for Energy (NICE), School of Energy and Power Engineering, Huazhong University of Science and Technology, Wuhan 430074, China

3 School of Optical and Electronic Information, Huazhong University of Science and Technology, Wuhan 430074, China

4 Mechanical Engineering Department, Faculty of Engineering, Kafrelsheikh University, Kafrelsheikh, Egypt

5 Mechanical Power Engineering Department, Faculty of Engineering, Tanta University, Tanta, Egypt

*Corresponding email: nuo@hust.edu.cn

# Abstract

Solar evaporation is important for many applications such as desalination, power generation and industrial drying. Recently, some studies on evaporation reported obtaining high energy efficiency and evaporation rate, which are based on floating evaporation setup (FES) with nanomaterials. Here, we proposed a new cheap and simple setup, named as airing evaporation setup (AES). It shows that the energy efficiency of AES reaches up to 87 % under 1 kW/m$^2$ of solar irradiation, which is 14% higher than that of FES. Meanwhile, the total evaporation rate of AES is about 20% higher than that of FES. The theoretical analysis reveals that the main reason for a better performance of AES is the increasing evaporation area. More interesting, AES could be used for designing portable systems due to its simplicity and flexibility. Furthermore, we show that AES and the corresponding wick material can be used in solar desalination, textile quick-drying and warm-keeping.

1. **Introduction**

**Enhancing the energy efficiency and evaporation rate of solar evaporation has attracted great attention during the past decades** due to its immense utility in many fields such as water purification or desalination [1-3], power generation [4], phase change cooling [5], industrial drying [6, 7], and so on. There are many traditional ways to enhance solar evaporation, such as using dye, solar collectors or black plate to get high solar energy absorption and high water evaporation rate [8, 9]. On the other hand, nanotechnology shows a better potential in improving solar evaporation. **Therefore, the solar absorber material and system design based on nanotechnology becomes a hot topic during the past decade** [10, 11].

**Nanofluid evaporation setup (NES) is one of the strategies for improving solar evaporation.** Researchers found that dispersing metallic nanoparticles into water can significantly enhance the solar energy absorption and heat transfer, which lead to significant increase in the evaporation rate [12-16]. However, the high evaporation rate requires extremely high irradiation density (>100 kW/m$^2$), which is far from practical application due to its high cost. Later, some researchers turned to carbon-based nanofluid [17-19], and they found that the energy efficiency can reach up to 70% under only one sun (1kW/m$^2$ of solar irradiation) [19]. However, the stability of nanofluid remains a challenge, which limits the application of nanofluid in solar evaporation [20].

**Another more effective strategy is utilizing floating evaporation setup (FES).** In FES, solar energy is absorbed on the top surface of floatable materials where creates heat localization[21]. The floatable materials are floating particles [22, 23], foams [21, 24-26], or films [27-30], which usually are hydrophilic and have high solar absorptivity. Low thermal conductivity is also required in order to localize heat at the top surface. The energy efficiency of evaporation reaches up to more than 80% under one sun, when using the gold nanoparticles or structures coated foams [31], nano porous membrane [32], carbon aerogels [26, 33-35], or carbonized biomaterials [36-38]. It is reported that the energy efficiency may reach up to 94% under one sun by using hierarchically nanostructured gels based on polyvinyl alcohol and polypyrrole [39]. Nevertheless, most of the floatable nanomaterials for high efficiency evaporation are complex and costly, which limits its application widely.

**Herein, we propose a cheap, simple and portable solar evaporation setup, named airing evaporation setup (AES), which also has high energy efficiency and evaporation rate.** Firstly, to verify the better performance of AES, the energy efficiency and evaporation rate of AES are measured and compared with FES and NES. Besides, the dependency of energy efficiency on the width of wings and concentration of particles are also studied. Secondly, the theoretical analysis is established to understand the mechanism of the high efficiency of AES. Finally, the potential applications of AES in solar desalination, textile quick-drying and warm-keeping are illustrated and discussed.

## 2. Experimental setup

**Figure 1 shows the schematic diagrams of the experimental setup of both AES and FES.** As shown in Fig. 1a, the wick material hangs above the water in AES.

Here, linen is chosen as the wick material due to its excellent capillary action [40, 41]. The shape of linen is circular with two wings which are partly immersed in water (Fig. 1b). Through the wings, water is absorbed up into the circular linen where is used to absorb solar energy and evaporate water. To study the effect of wings, wings with different width (W) were designed, while the diameter (D) is kept as the same (5 cm). The characteristics of linen are shown in Table 1. To enhance the solar absorption and heat transfer, carbon black nanoparticles are dispersed on the surface of linen. The size of nanoparticles is around 40 nm, as measured by transmission electron microscopy (Fig. 2).

**To compare with AES, the floating evaporation setup(FES), which uses floatable materials, is also designed according to the references** (Fig. 1c) [21, 42-45]. In FES, all the other surfaces of wick material are insulated with EPE (Expandable Polyethylene) foam, except the top surface which is the only surface for evaporation.

**The schematic diagram of AES shows that there is no insulation foam, that is simpler than FES.** The wick material in AES can be supported by wires or strings which have negligible volume and cost. The flexible wick material can be easily folded or rolled, and so, it is very convenient for transportation and storage compared to the thick foam in FES [11, 46] and unstable nanofluids [47, 48]. Therefore, AES is more suitable for making portable systems, such as portable solar still. Removing the foams also decreases the cost, because of the foams account for around 1/3 to 1/2 of the total cost in FES, according to the retail price [45].

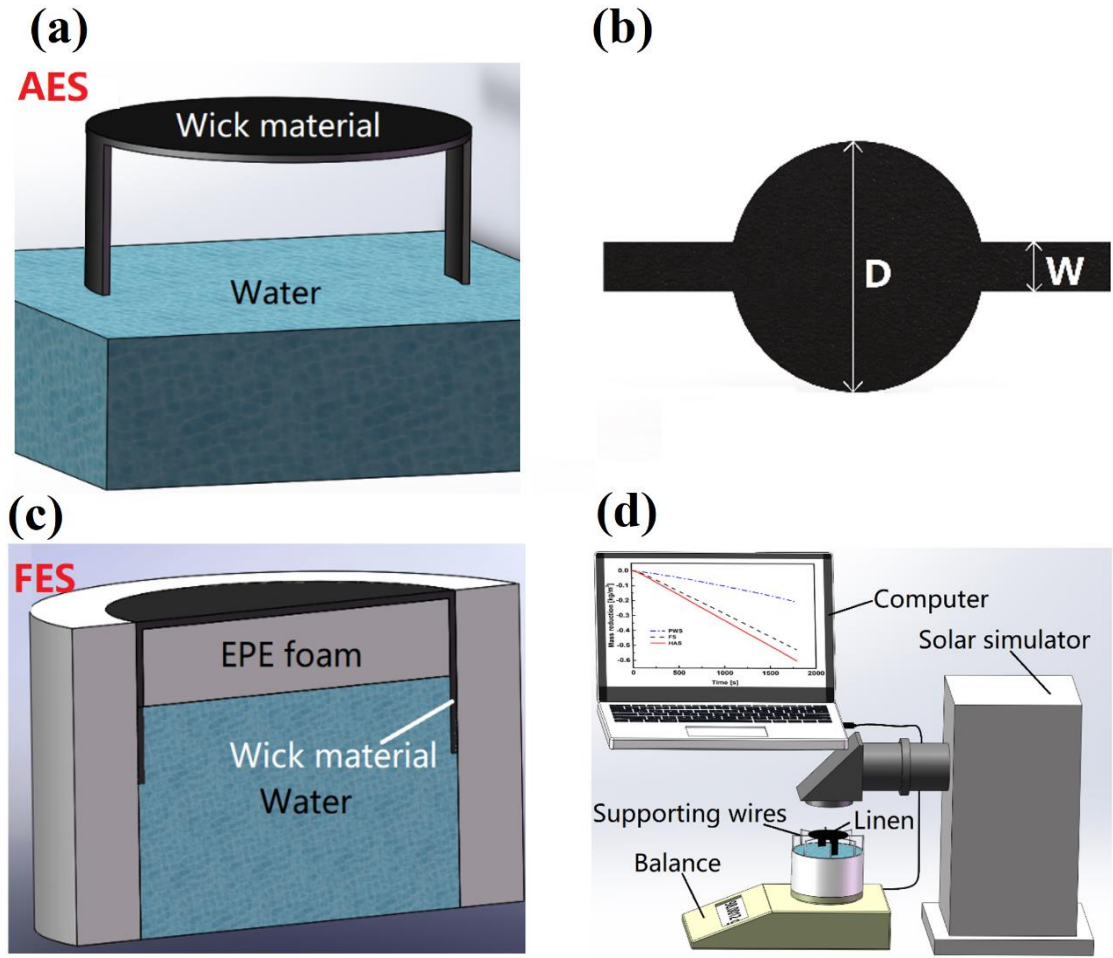

**Figure 1.** (a) The schematic diagram of AES. The wick material hangs 2 cm above the water with two wings partly immersed in water. (b) The structure of the wick material. The diameter of the circle for absorbing solar energy is fixed at 5 cm. the width of wings varies from 0.4 cm to 3 cm. (c) The schematic diagram of FES. (d) The experiment setup of the evaporation measurement. The surface of the bulk water is covered by a layer of cling film. The wick material hangs above the water supported by wires.

**The measurement setup for AES and FES is shown in Fig. 1d.** A solar simulator (CEL-S500, AM1.5 filter) was used to generate solar beam. The solar intensity was adjusted to 1 kW/m$^2$ by using a power meter (PM-150-50C). The mass reduction during the evaporation process was measured by an electric balance (Sartorius Practum 224), the data were recorded by a computer via a USB cable. The ambient temperature and humidity during the experiment were controlled at around 24 °C and 50%, respectively.

**Table 1 The characteristics of the wick material.**

| material | characteristic | value |
| --- | --- | --- |
| linen | Rate of moisture regain, (%) | 12.5 |
|  | Mass density, (g/m$^2$) | 250 |
|  | Thickness, (mm) | 1 |

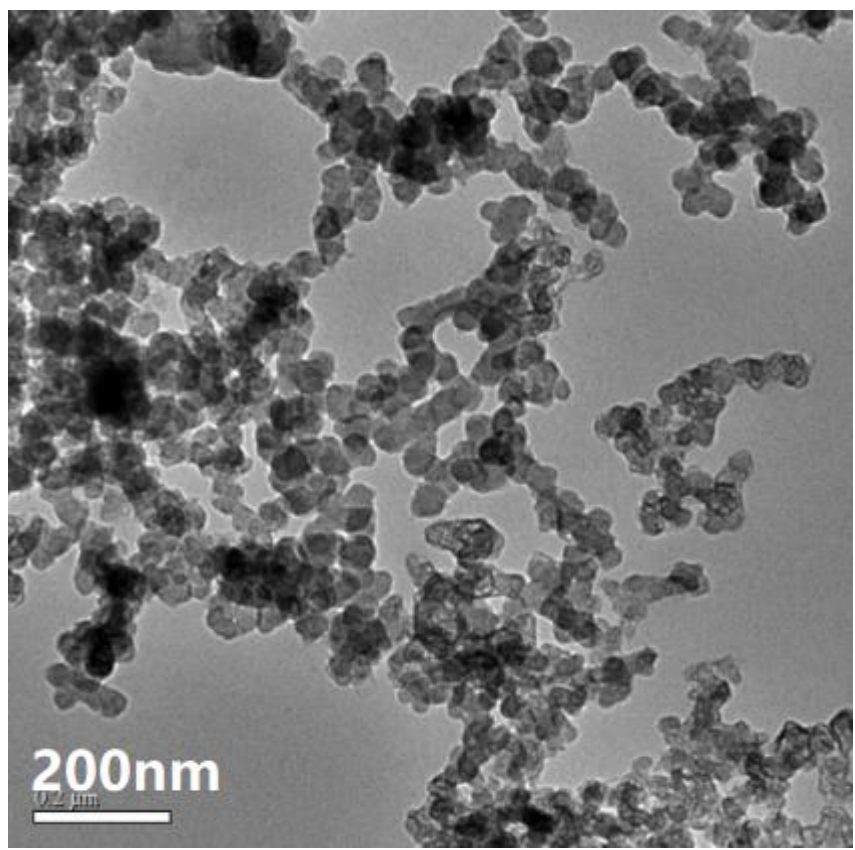

**Figure 2.** The TEM image of the carbon black nanoparticles. The scale bar is 200 nm.

## 3. Results and discussions

**The amount of water evaporated under one sun (1 kW/m² of solar irradiation) was measured as the mass reduction of systems.** Fig. 3a shows the mass reduction along time of three different solar evaporation setups, traditional pure water evaporation setup (PWES), NES, FES and AES. It should be noted that the values of natural evaporation (the evaporation under dark environment) are removed in all cases, which are around 0.07, 0.07, 0.09, 0.2 kg/(m²·h) for PWES, NES, FES and AES, respectively. The reduction rate of AES is the highest, which is about 10% , 115% and 200% higher than FES, NES and PWES, respectively. This means that AES not only simplifies the structure, but also increases the evaporation rate, compared to the commonly used effective FES.

**To further enhance the evaporation rate and energy efficiency, the structure of wick material in AES is optimized by changing width of wings.** The energy efficiency as a function of width of wings is shown in Fig. 3b. The energy efficiency, $\eta$, is defined as [21]:

$$\eta = \frac{\Delta m \cdot h_{LV}}{Q} \tag{1}$$

where $\Delta m$ is the evaporation rate, i.e. the hourly mass reduction per unit area; $h_{LV}$ is the total enthalpy of phase change (~2460 kJ/kg), which contains latent heat and sensible heat; $Q$ is the nominal direct solar irradiation, which is fixed at 1 kW/m². In AES, the efficiency is around 76% when the width of wings is 0.4 cm. Then the efficiency increases when the width of wings increases, and finally the efficiency converged to 87% when the width of wings is above 2 cm. In comparison, the efficiency is only around 28%, 40% and 73% for PWES, NES and FES, respectively.

**In addition to width of wings, the mass concentration of particles on wick material also affect evaporation.** The energy efficiency of AES is only 54% when there are no carbon black nanoparticles, which is 30% lower than that of when the concentration of particle is 15 g/m$^2$ (Fig. 3d). It should be noticed that the efficiency increases dramatically when the mass concentration of particles increases from 0 to 1.25g/m$^2$. However, the efficiency only slightly increases for concentration of particles further increases to 15 g/m$^2$. Therefore, only a very low concentration (>1.25 g/m$^2$) is required for keeping a high efficiency, which means that the cost of nanoparticles is less than $ 0.2/m$^2$ according to the retail price of carbon black nanoparticles. The extremely low cost of nanomaterials is very promising in practical applications.

**Besides the above mentioned evaporation rate induced by solar, natural evaporation rate is also very important in evaporation system.** Fig. 3c shows that AES also gives much higher natural evaporation rate than that of FES, which indicates that AES can absorb more ambient energy for water evaporation. For the width of wings at 3 cm, the water evaporated around 0.12 kg/m$^2$ in 30 min, which is 150% higher than that of FES. The results also demonstrate that the natural evaporation is an increasing function of the width of wings, due to the increased evaporation area. The total evaporation rate (natural evaporation + solar induced evaporation) of AES is around 20% higher than that of FES.

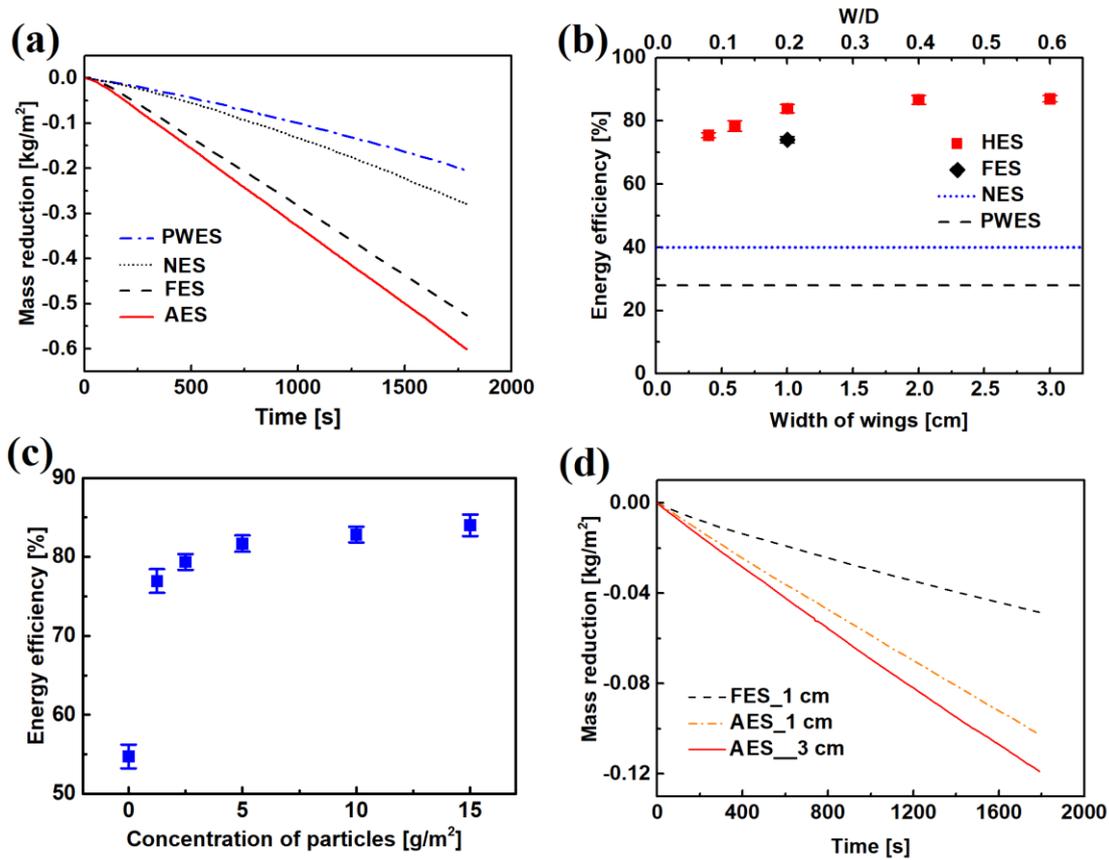

**Figure 3** (a) The mass reduction along time for PWES, NES, FES and AES for water depth at 5 cm. The concentration of carbon black nanoparticles in NES is 0.5wt.% according to Ref. 16 . The width of wings is 1 cm and the concentration of particles is 10 g/m$^2$. (b) The effect of the width of wings on energy efficiency. The concentration of particles is 10 g/m$^2$. (c) The effect of the concentration of particles on energy efficiency. The width of wings is fixed at 1 cm. (d) The mass reduction of natural evaporation for FES and AES with different width of wings. The concentration of particles is 10 g/m$^2$.

**To understand why AES is better than FES on theory, two setups have been analyzed based on the classic evaporation theory** [49, 50]:

$$\dot{m} = \varepsilon(P_s - P_v)\sqrt{\frac{M}{2\pi RT}} \qquad (2)$$

$$P_s = e^{\left(25.317 - \frac{5144}{T}\right)} \qquad (3)$$

where $\dot{m}$ is the evaporation rate of water at temperature $T$, $\varepsilon$ is the evaporation coefficient, which is a constant. $P_s$ is the saturated vapor pressure of water at temperature $T$. $P_v$ is the vapor pressure of the ambient, which is 1486 Pa according to the ambient temperature and humidity. $M$ is the relative molecular mass of water, $R$ is the universal gas constant (8.314 J·mol⁻¹·K⁻¹).

In AES, the structure of wick material will affect the evaporation. As shown in Fig. 4a, the evaporation surface can be divided into 2 parts, the outer surface and inner surface. For the outer surface, the evaporation rate, $\dot{m}_{out}$, is obtained by Eq. (2) and (3) as:

$$\dot{m}_{out} = 0.587\varepsilon \left(e^{\left(25.317 - \frac{5144}{T}\right)} - 1486\right) T^{-0.5} \qquad (4)$$

However, for the inner surface, the vapor diffusion is blocked by the wings, thus the vapor from the inner surface is more difficult to diffuse into the ambient compared to the outer surface. With the width of the wings increases, the diffusion area decreases, hence the less evaporation rate per unit evaporation area. When wings fully cover the inner surface, i.e., the total width of wings (2W) equals to the perimeter of the circle ($\pi D$), vapor diffusion will be completely blocked and evaporation rate will be zero, which is similar to the covered surface in FES (Fig. 4b). Therefore, the evaporation rate from the inner surface, $\dot{m}_{in}$, is defined as:

$$\dot{m}_{in} = 0.587\varepsilon \left(1 - \frac{2W}{\pi D}\right)\left(e^{\left(25.317 - \frac{5144}{T}\right)} - 1486\right) T^{-0.5} \qquad (5)$$

Thereby, the overall mass flux of evaporation in AES, $\dot{M}$, which is the integral

of $\dot{m}$, is given by:

$$\dot{M} = \int \dot{m}_{out} \Delta A_{out} dA_{out} + \int \dot{m}_{in} \Delta A_{in} dA_{in} \qquad (6)$$

where $A_{out}$ and $A_{in}$ are the evaporation area of the outer and inner surface, respectively.

Herein, we obtained the temperature distribution of water by using COMSOL based on energy conservation:

$$Q_{sol} = Q_{rad}(T) + Q_{cov}(T) + Q_{eva}(T) + Q_{cod}(T) \qquad (7)$$

where $Q_{sol}$ is the energy of solar irradiation, $Q_{rad}(T)$, $Q_{cov}(T)$ and $Q_{cod}(T)$ are the radiation, convection and conduction loss, respectively, which are determined by the ambient and water temperature. $Q_{eva}(T)$ is the energy carried away by vapor based on Eq. (4) and Eq. (5). Herein, the slow mass transfer of water in the wick material is ignored, which has negligible effect to the overall heat transfer process (<2%). $Q_{eva}(T)$ is calculated as the following:

$$Q_{eva}(T) = \dot{m} h_{fg} \qquad (8)$$

where $h_{fg}$ is the latent heat of phase change.

**The theoretical temperature distribution of water in AES and FES are shown in Fig. 4a and 4b, respectively.** The water temperature in AES reaches up to 307 K which is 8 K lower than that of FES. AES has more evaporation area which makes the water unable to reach high temperature. The water temperature at the wings in AES is the same as or even lower than the ambient temperature. This is due to the water evaporation at the wings carries a lot of heat away. However, the water temperature at the wings in FES is much higher than the ambient temperature, thus more heat is transferred to the bulk water compared to AES. Meanwhile, the high surface temperature in FES also increases the convection and radiation heat loss. Therefore, the efficiency of FES is lower than AES. The results imply that for high efficiency evaporation, creating more evaporation surface might be more important than creating high temperature.

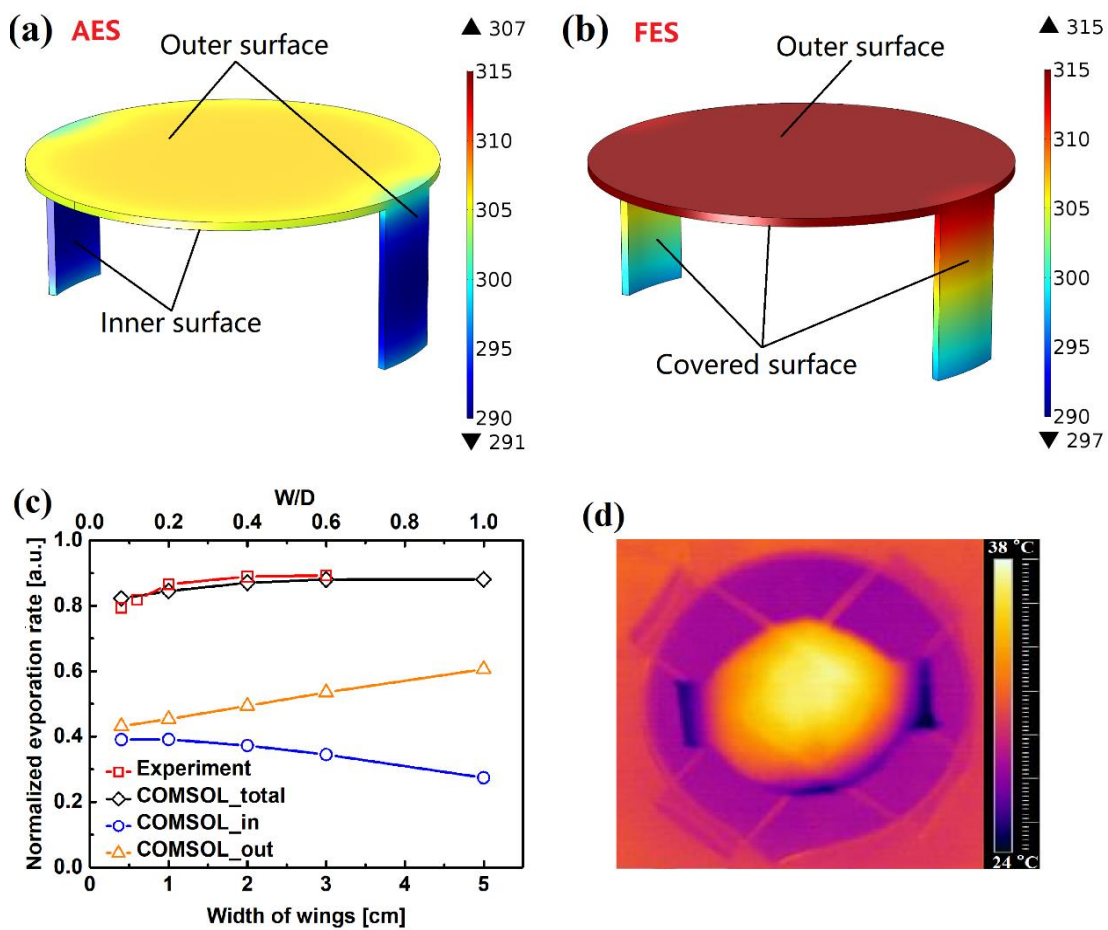

**Figure 4** (a) The water temperature in AES for width of wings at 1 cm. The outer (inner) surface indicates that the vapor diffusion near the surface isn't (is) blocked by the wings. (b) The temperature distribution of water in FES for the width of wings at 1 cm. The covered surface means the surface is covered by insulating material. (c) The normalized evaporation rate of AES. "Total" indicates the total evaporation rate, "in" and "out" means the evaporation rate from the inner and outer surface, respectively. In simulation, the water temperature at the end of the wings is 297.15 K, which is the same as the temperature of the bulk water. The inlet heat flux on the top surface was fixed at 1 kW/m$^2$. (d) Thermal image of HAS system, the hottest region is the wick material.

**Based on the water temperature, theoretical evaporation rate can be obtained by Eq. (4)-(6).** Herein the evaporation rate includes both natural evaporation and solar induced evaporation. Fig. 4c shows that the evaporation rate of the outer surface in AES increase linearly with the width of wings. On the other side, the evaporation rate of the inner surface decreases when the width of wings increases due to the less area for vapor diffusion as discussed in Eq. (5). Therefore, the total evaporation rate of AES increases first and then converges with the increasing of width of wings. The theoretical calculations matched well with the experimental data (Fig. 4c and 4d).

**Due to the high efficient evaporation and simple structure of AES as shown above, AES and the corresponding nano wick material have many potential applications,** such as solar desalination, textile quick-drying and warm-keeping. It is measured that the evaporation rate of salt water and fresh water is almost the same by using AES (Fig. 5a), which indicates that AES can be used in desalination efficiently. Besides, the high water evaporation rate of nanoparticles coated wick material also shows its potential in quick-drying textile. It is show that, when the moisture content on wick material is around 80%, the wick material with nanoparticles saves 40% of

drying time compared to that without nanoparticles (Fig. 5b). Moreover, due to the high solar absorptivity of carbon black nanoparticles [17, 43], the temperature of wick material with nanoparticles can be 30 °C higher than that of without nanoparticles under 1 kW/m² of irradiation (Fig. 5c). It means that nanoparticles coated textile could be used in some solar heating process, such as warm-keeping in winter days. **Therefore, this work gives more possibility in solar energy utilization and multifunctional textile designing.**

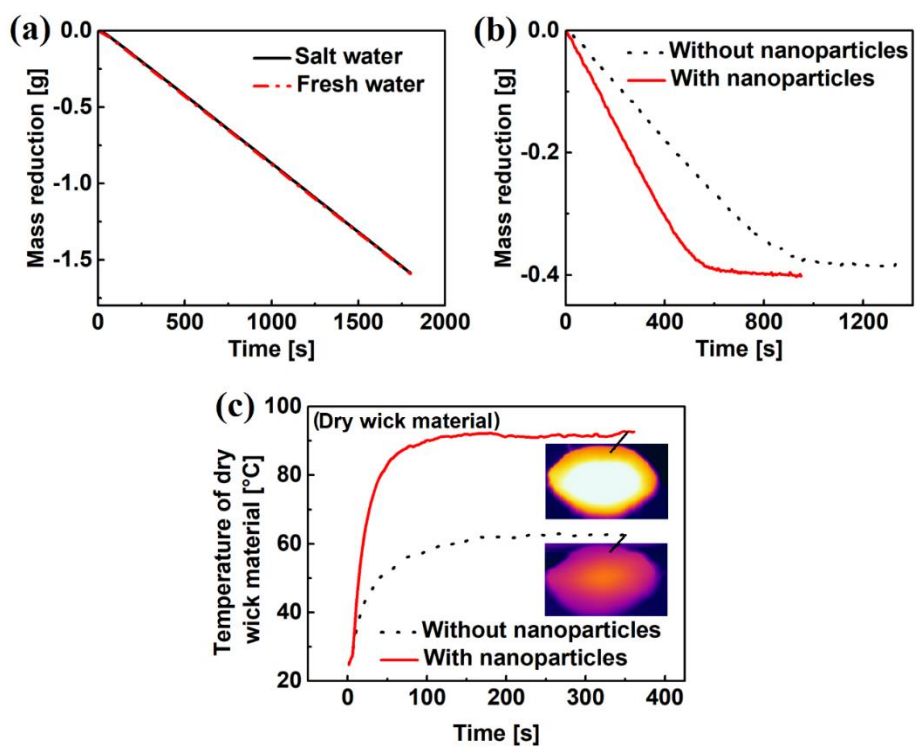

Fig. 5 (a) The mass reduction of AES system by using fresh water and salt water (3.5 wt.% of salt) (b) The mass reduction of drying process by hang-airing wet wick material (5 cm in diameter, 80% in moisture content) under 1 kW/m² of irradiation. (c) The temperature of different dry wick material under 1 kW/m² of irradiation. The insert figures are the thermal images of different wick material.

## 4. Conclusion

In conclusion, an airing evaporation setup (AES) has been proposed for high efficient solar evaporation by using a multifunctional nano wick material. The results show that the solar induced evaporation rate of AES is obtained around 14%, 120%, and 210% higher than that of floating evaporation setup (FES), nanofluid evaporation setup (NES) and traditional pure water evaporation setup (PWES), respectively. Moreover, it is found that the energy efficiency of AES depends on the width of wings and the nanoparticles concentration. The energy efficiency increases from 76 % to 87% when the width changes from 0.4 cm to 3 cm. When using a very low concentration (1.25g/m$^2$) of carbon black nanoparticles, the efficiency can also be increased as high as 20% compared to that without nanoparticles. The advantage of a low concentration makes a low-cost of material.

Besides the solar induced evaporation rate, AES also shows a high natural evaporation rate due to more evaporation area. Thus, the total evaporation rate of AES is around 20% higher than that of FES. The theoretical results predict that the increase of evaporation area of the inner surface in AES is the main reason of high efficient evaporation. However, it's found that wider wings block vapor diffusion from the inner surface, which limits the increasing of evaporation rate.

In the end, we show a discussion on the potential application of AES and the corresponding nano wick material in in designing multifunctional textile and solar energy utilization.

## 5. Conflicts of interest

There are no conflicts of interest to declare.

## 6. Acknowledgement

N.Y. was sponsored by National Natural Science Foundation of China (No. 51576076 and No. 51711540031), Natural Science Foundation of Hubei Province (2017CFA046) and Fundamental Research Funds for the Central Universities (2016YXZD006). The authors thank the National Supercomputing Center in Tianjin (NSCC-TJ) and China Scientific Computing Grid (ScGrid) for providing assistance in computations.